\documentclass[floats,floatfix,showpacs,amssymb,prd,twocolumn,superscriptaddress,nofootinbib,nolongbibliography,reprint]{revtex4-2}

\usepackage{amssymb,amsmath,verbatim,mathtools,needspace,enumitem,etoolbox,graphicx,physics,microtype,afterpage,bigints,gensymb,tabularx,xspace}

\usepackage{enumitem}
\setlist[itemize]{
  topsep=3pt,
  itemsep=1.5pt,
  parsep=1.5pt,
  partopsep=1.5pt
}

\usepackage[dvipsnames, usenames]{xcolor}
\definecolor{linkcolor}{rgb}{0.0,0.3,0.5}
\definecolor{dodgerblue}{HTML}{1E90FF}
\usepackage[unicode, colorlinks=true, linkcolor=linkcolor, citecolor=linkcolor, filecolor=linkcolor,urlcolor=linkcolor, pdfusetitle]{hyperref}
\usepackage[all]{hypcap}
\usepackage[T1]{fontenc}
\usepackage[utf8]{inputenc}
\usepackage{orcidlink}

\newcommand{\ssim}{\mathchar"5218\relax\,}

\usepackage{seqsplit}

\makeatletter
\newcommand*{\balancecolsandclearpage}{\close@column@grid \cleardoublepage \twocolumngrid}
\makeatother

\newcommand{\milan}{\affiliation{Dipartimento di Fisica ``G. Occhialini'', Universit\'a degli Studi di Milano-Bicocca, Piazza della Scienza 3, 20126 Milano, Italy}}
\newcommand{\infn}{\affiliation{INFN, Sezione di Milano-Bicocca, Piazza della Scienza 3, 20126 Milano, Italy}}
\newcommand{\nottingham}{\affiliation{Nottingham Centre of Gravity \& School of Mathematical Sciences, University of Nottingham, University Park, Nottingham NG7 2RD, United Kingdom}}

\renewcommand{\section}[1]{\textbf{\textit{#1}} ---}

\begin{document}

\title{Variance of gravitational-wave populations}

\author{Alessia Corelli$\,$\orcidlink{0009-0003-8157-243X}}
\email{a.corelli1@campus.unimib.it}
\milan

\author{Davide Gerosa$\,$\orcidlink{0000-0002-0933-3579}}
\milan \infn

\author{Matthew Mould\,\orcidlink{0000-0001-5460-2910}}
\nottingham

\author{Cecilia Maria Fabbri\,\orcidlink{0000-0001-9453-4836}}
\nottingham


\date{\today}

\begin{abstract}

We quantify the impact of finite catalog size, or ``catalog variance,'' on current gravitational-wave population analyses. The distribution of merging binary black holes is commonly reconstructed via hierarchical Bayesian inference, with uncertainties reported as credible intervals. Such intervals are conditioned on the specific realization of the observed events and are therefore themselves subject to variability arising from the finite size of the catalog. We estimate this ``uncertainty on the uncertainty'' using statistical bootstrapping applied to 
 data segments containing both detected events and sensitivity injections. 
Applying this framework to GWTC-4, we find that the inferred population distributions exhibit substantially broader uncertainties than those obtained in a standard single-catalog analysis. In particular,
the $\ssim 35\,M_\odot$ peak in the primary-mass distribution %
is largely absorbed by statistical fluctuations once catalog variance is taken into account.
Unlike other studies that rely on simulating catalogs by assuming an underlying population,
this work provides the first
data-driven assessment of the uncertainty intrinsic to the observed gravitational-wave catalog.
Accounting for catalog variance is important for drawing robust astrophysical conclusions from gravitational-wave data, avoiding inferences driven by a particular finite realization rather than genuine population features.

\end{abstract}

\maketitle

\section{Catalog variance} %
Population fits are among the primary scientific outcomes of gravitational-wave (GW) observing campaigns~\cite{2019ApJ...882L..24A,2021ApJ...913L...7A,2023PhRvX..13a1048A,2025arXiv250818083T}. Their goal is to characterize the distribution of observed events while accounting for measurement errors and selection effects~\cite{2024arXiv241019145C}. The most recent study based on the GWTC-4 catalog~\cite{2025arXiv250818083T} %
by the LIGO--Virgo--KAGRA interferometers~\cite{2015CQGra..32g4001L, 2015CQGra..32b4001A, 2021PTEP.2021eA101A} includes data from 153 binary black hole (BH) coalescences. As the detectors continue to collect data, the available dataset grows, enabling an increasingly precise characterization of the BH binary population. Nevertheless, we remain fundamentally limited to a single realization of the detection catalog. By construction, {there exists only one catalog containing all the events we have observed}.

This effect, which we refer to as ``catalog variance,'' is somewhat similar to the so-called ``cosmic variance'' quoted, for example, in analyses of the cosmic microwave background, where the power spectrum at low-$\ell$ multipoles is fundamentally limited by the fact that only one sky is observable.
There is, however, a crucial difference, as the catalog variance considered here stems from the finite number of detections at our disposal~\cite{2024PhRvD.109h1302P}.

Quantifying the impact of the catalog variance on current GW results is crucial for a robust astrophysical interpretation of the data. Are the features observed in the  BH binary population genuine astrophysical properties of BHs, or are they statistical fluctuations arising from our finite set of observations? To this end, the most commonly adopted strategy is to perform posterior-predictive checks using synthethic observations~\cite{2022PASA...39...25R}. In this framework, one simulates multiple catalog realizations assuming a population that does not exhibit the feature of interest, and then checks how often such a feature is nevertheless inferred.
This approach has been used to test the presence of a peak in the BH mass spectrum~\cite{2023ApJ...955..107F}, the putative overabundance of BHs with aligned spins~\cite{2025PhRvD.112h3015V}, the presence of a mass gap~\cite{2026arXiv260211282M}, and possible mass–spin correlations~\cite{2021ApJ...922L...5C}.
These approaches simulate catalogs rather than operate directly on the observed data and therefore require specific assumptions about the underlying population. As a consequence, they do not quantify the catalog variance associated with the actual realization of the observed catalog, but only assess the variance across mock catalogs constructed to resemble it.

In this work, we present the first data-driven estimate of the catalog variance in GW astronomy.
{Our strategy is based on statistical bootstrapping of the observed data stream itself, which we apply to GW population fits. Crucially, this targets the variance of the actual catalog of events that was observed rather than that of simulated catalogs.}

\begin{figure*}
    \centering
    \includegraphics[width=0.85\textwidth]{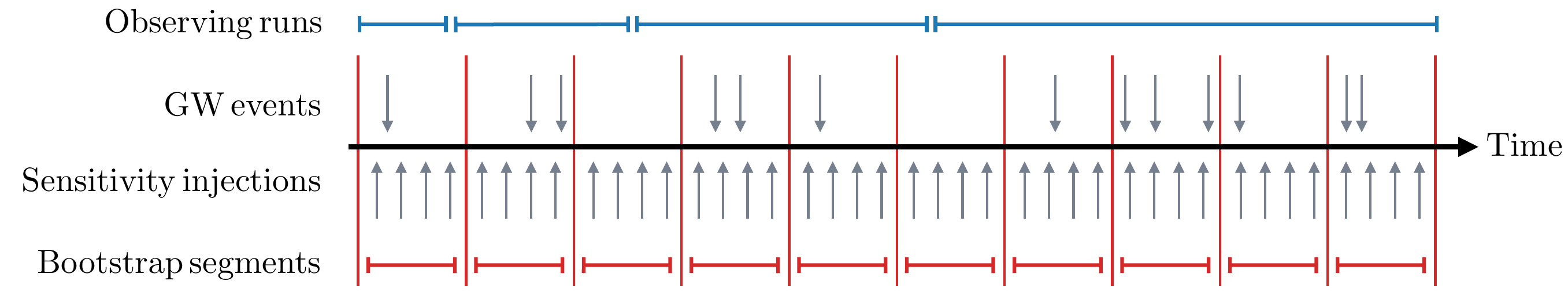}
    \caption{%
    Schematic representation of our resampling strategy. Detected events and sensitivity injections (grey arrows) are placed on a contiguous timeline, excising the breaks between observing runs (blue segments at the top). This timeline is then divided into $N$ equal-length segments (red, at the bottom), which are resampled with repetition to estimate the variance of the resulting population analysis.
    }
    \label{fig:timeline}
\end{figure*}

\section{Resampling}
\label{sec:Variance via resampling}
Bootstrapping is a widely used statistical procedure for computing the distribution of an estimator by resampling data with replacement~\cite{tibshirani1993introduction,davison1997bootstrap} (an alternative approach based on leave-one-out methods is presented in the End Matter; see, e.g., Ref.~\cite{2013acna.conf...15L} for caveats on resampling techniques). In this paper, we design an application of bootstrapping to hierarchical Bayesian analyses affected by selection biases~\cite{2019MNRAS.486.1086M,2022hgwa.bookE..45V}. Our problem is that of estimating the posterior distribution $p(\lambda | d)$, where $\lambda$ are the population parameters and $d$ are the data. Once samples $\lambda\sim p(\lambda | d)$  are obtained, these can be used to reconstruct the posterior of the differental merger rate $d \mathcal{R}/d\theta = \mathcal{R} \,p(\theta|\lambda)$, where $\theta$ are the single-event parameters (in our case the masses, spins, and redshifts of binary BH mergers).

Crucially, the data $d$ here encode the entire LVK data stream, not just the short stretches of data containing a GW event. This is important, as the time when the detectors have \emph{not} seen a GW event enters the determination of selection effects.
In practice, the {observational} ingredients entering GW population fits are (i)
{the likelihood of each observed BH binary represented by posterior samples~\cite{2025arXiv251011197A}}
and (ii)
{the survey sensitivity represented by a set of simulated signals injected into search pipelines~\cite{2025PhRvD.112j2001E}}.
Bootstrapping over the data $d$ requires designing a resampling approach for both these ingredients.

Our strategy is that of resampling the observing time. We place all detected events and sensitivity injections on a single timeline and divide this into $N$ equal segments. A schematic representation is reported in Fig.~\ref{fig:timeline}.
These time segments are then resampled with repetition, thus creating a new dataset $d^*$
used to estimate $p(\lambda | d^*)$.
This procedure is repeated many times, and the
{variation between the resulting $p(\lambda | d^*)$ across realizations $d^*$}
is interpreted as the variance of the posterior $p(\lambda | d)$ obtained with the true dataset. The same applies to the reconstructed rate~$d \mathcal{R}/d\theta$.

Our setup mirrors the default ``strongly modeled approach'' from the BH-binary population analysis of Ref.~\cite{2025arXiv250818083T} (see their Table~1).
We consider the 153 BH binary events detected up to GWTC-4; single-event posterior samples are provided at Refs.~\cite{ligo_scientific_collaboration_and_virgo_2022_5117702,ligo_scientific_collaboration_and_virgo_2023_5546662,ligo_scientific_collaboration_and_virgo_2025_16053483}. Each event is labeled with the time of peak GW amplitude, which is the same time reported by convention in the name of the event itself~\cite{2025ApJ...995L..18A}. We use the same set of injections~\cite{ligo_scientific_collaboration_2025_16740128} used to estimate sensitivity as in Ref.~\cite{2025arXiv250818083T}
and label each entry according to the reported geocenter time.

We exclude periods during which the detectors were not operational. We consider the time intervals between consecutive injections and identify four long gaps, %
corresponding to breaks between the LVK observing runs.
These gaps are excised and substituted with the median separation between injections (23\,s), resulting in a timeline of 2.616 years. The same time transformation is then applied to the detected events.

For the results presented in the main body of the paper, we consider $N=150$ time segments---a number chosen to be comparable to the number of detected events. In the End Matter, we show that our results are largely unchanged even if $N$ varies by up to two orders of magnitude. Each event or injection is uniquely assigned to one of the $N$ time segments, neglecting possible splitting between two adjacent segments. This is a good approximation because the length of each time segment (a few months) is much larger than the duration of a BH binary transient (a few seconds).

We perform a total of 700 %
bootstrap realizations, repeating the full population analysis each time. The chance of extracting the same bootstrap dataset $d^*$ twice is vanishingly small: the number of possible resamplings is $\binom{2N-1}{N}$, which for $N=150$ is $\sim 5\times 10^{88}$.

When estimating detectability, we use the same thresholds in the signal-to-noise ratio (SNR) $> 10$ (for the O1 and O2 observing runs) and false-alarm rate (FAR) $< 1 \,\mathrm{yr}^{-1}$ (for O3 and O4), and the same total analysis time of 2.12 yr as in Ref.~\cite{2025arXiv250818083T}. This is shorter than the length of the timeline discussed above because it accounts for additional periods of detector downtime that were not excised; those data-taking breaks are bootstrapped together with the injections. %
When the injection campaign was performed~\cite{2025PhRvD.112j2001E}, some low-SNR systems were automatically flagged as non-detectable to save computational time. The available dataset~\cite{ligo_scientific_collaboration_2025_16740128} reports the total number of these ``hopelessly quiet'' injections, but does not list their parameters and times. The former is all that is needed for a population analysis after all, but for the present application this implies that the ``hopelessly quiet'' injections cannot be bootstrapped. As a workaround, we bootstrap only the synthetic signals that were actually injected into noise, while keeping the number of the ``hopelessly quiet'' ones fixed to that of the original dataset. There are
$\sim10^9$
of these ``hopelessly quiet'' injections; assuming a Poisson counting process, fixing this number introduces a spurious effect at the
$\sqrt{10^{-9}} \simeq 3\times 10^{-5}$
level.

\begin{figure*}[!p]
    \centering
    \includegraphics[width=0.97\textwidth]{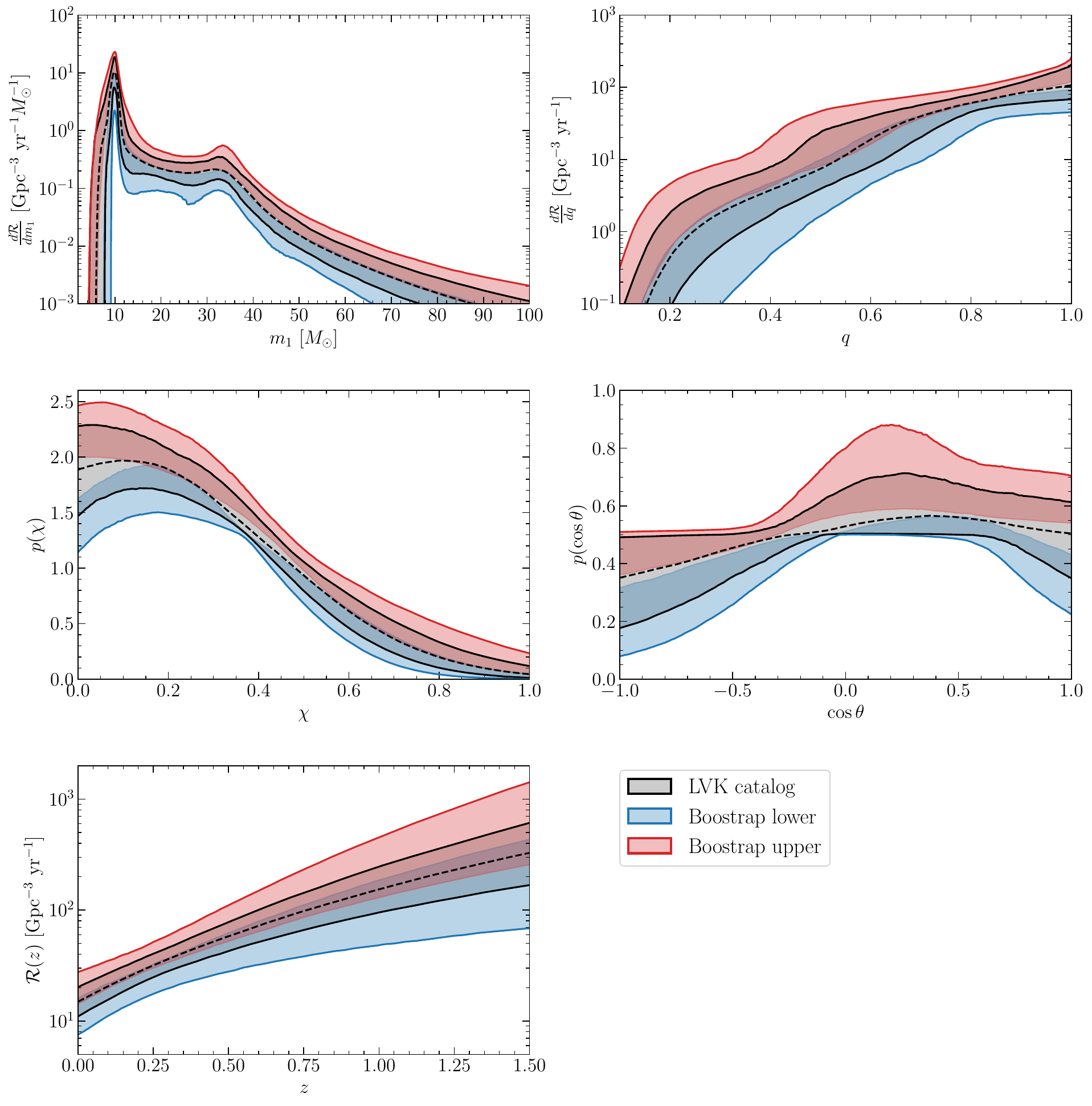}
    \caption{Reconstructed population distributions for BH binary parameters, showing the effect of catalog variance. The black dotted and solid curves, along with the shaded areas, represent median and 90\% credibility regions obtained from the GWTC-4 data alone, reproducing the results of Ref.~\cite{2025arXiv250818083T}. Red and blue shaded bands indicate the variability of the upper and lower bounds of these credibility regions across 700 bootstrap realizations of the catalog. In particular, red (blue) solid curves indicate the upper (lower) 90\% confidence interval on the upper (lower) 90\% credibility interval from GWTC-4, providing an estimate of the uncertainty associated with performing the population analysis on a single finite dataset. From left to right and from top to bottom, we show the differential merger rate as a function of primary mass $m_1$ at $z=0.2$, the differential merger rate as a function of mass ratio $q$ at $z=0.2$, the (normalized) distribution of spin magnitudes $\chi$, the (normalized) distribution of the cosine of the spin tilts $\theta$, and the merger rate as a function of redshift $z$ (these choices were made to facilitate direct comparison with Ref.~\cite{2025arXiv250818083T}).
    }
    \label{fig:catalog_variance}
\end{figure*}

\section{Current data}
Figure~\ref{fig:catalog_variance} shows the reconstructed populations in primary source-frame mass $m_1$, mass ratio $q$, spin magnitude $\chi$, spin direction $\theta$, and redshift $z$. The black shaded area represents the 90\% credibility region obtained from the GWTC-4 data and fully reproduces the results of Ref.~\cite{2025arXiv250818083T}. This region is derived from a single hierarchical analysis of the real dataset.

When repeating the inference using bootstraps, each iteration yields a different reconstructed population and, correspondingly, a different pair of upper and lower bounds for the 90\% credibility interval at each value of the parameter shown on the horizontal axis. At fixed parameter value (e.g., fixed $m_1$ or $z$), we therefore obtain a distribution of upper bounds and a distribution of lower bounds across the 700 bootstrap realizations. The red and blue shaded areas in Fig.~\ref{fig:catalog_variance} summarize these distributions: the red region encloses 90\% of the bootstrapped upper credibility bounds, while the blue region encloses 90\% of the bootstrapped lower credibility bounds. These bands quantify the variability of the inferred 90\% credibility region under resampling of the dataset. We interpret them as ``error bars'' on the black shaded region, i.e., as an estimate of the uncertainty associated with performing the population analysis on the single, finite GWTC-4 dataset.

The main result of this analysis is a significantly wider uncertainty on the inferred population distributions. In other words, while the commonly quoted 90\% credibility region extends between the two solid black curves, accounting for the catalog variance increases the ``effective uncertainty'' to the region enclosed by the solid blue and solid red curves.
\begin{itemize}
\item
For the primary mass, the broader uncertainty induced by catalog variance has a noticeable impact: the mass spectrum becomes compatible with a single power law between $\ssim 15\,M_\odot$ and $\ssim 100\,M_\odot$, without requiring an additional peak at $\ssim 35\,M_\odot$. This statement is supported by the confidence intervals shown in the top-left panel of Fig.~\ref{fig:catalog_variance}, within which a straight line can be drawn across that mass range. It is further corroborated by visual inspection of the individual bootstrap realizations, many of which do not display a clear peak around $\ssim 35\, M_\odot$.
On the other hand, the main peak at $\ssim 10\,M_\odot$ \emph{cannot} be absorbed into a single power law, even after accounting for the increased uncertainty due to the catalog variance.

\item As for the mass ratio, accounting for the catalog variance increases the error bars by about an order of magnitude, with the distribution remaining well compatible with a single (tapered) power law.

\item A similar conclusion applies to the spin magnitudes, with the size of the error bars increasing while the overall shape of the distribution remains roughly unchanged. Notably, the excess of low-spin BHs ($\chi\lesssim 0.3$) appears to be a robust feature.

\item With the enlarged uncertainties induced by catalog variance, the spin-orientation distribution is largely compatible with a fully isotropic configuration (flat in $\cos\theta$). In particular, the uncertainties associated with the peak around $\cos\theta \sim 0.2$ nearly double. This should be regarded as a caution against overinterpreting current spin-orientation features, which may not be robust.

\item The redshift distribution is also more uncertain.
For instance, the local merger rate ($z=0$) lies between $\ssim 11$ and $\ssim 20$ Gpc$^{-3}$yr$^{-1}$ at 90\% credibility when considering the single realization provided by GWTC-4, but extends from $\ssim 7$ to $\ssim 28$ Gpc$^{-3}$yr$^{-1}$ when accounting for the catalog variance.
However, the merger rate is consistently found to increase with redshift up to $z\sim 1.5$.
\end{itemize}

\section{Interpretation}  
Uncertainty in GW population studies is usually quantified in terms of credible intervals, e.g., regions of parameter space that encompass the central 90\% of the posterior distribution. As quantified in this paper, such intervals are themselves subject to a fundamental uncertainty due to the finite size of the observed catalog. In short, the catalog variance can be interpreted as the ``uncertainty of the uncertainty'' in GW population fits.
Operationally, accounting for the catalog variance results in inflated error bars, hence to astrophysical constraints that are intentionally more conservative and therefore less restrictive than those reported elsewhere.

It is important to note that, at least within a Bayesian framework, the posterior reconstructed from a given dataset $d$ is the final result of the analysis, and the associated uncertainties are precisely what they should be when conditioning on $d$. Nonetheless, it remains a pertinent question to assess the robustness of the inferred features against the fact that the dataset $d$ has finite size.
To the best of our knowledge, the application presented here is the very first attempt to address this issue \emph{directly from the observed GW catalog itself}, rather than from simulated catalogs constructed to resemble it.

Our analysis indicates that the $m_1 \sim 35\,M_\odot$ peak in the primary mass distribution does not appear to be stable under catalog variance and may instead be consistent with a statistical fluctuation of the current dataset, even with the addition of GWTC-4 data. While previous LVK studies have reported high confidence in this feature~\cite{2021ApJ...913L...7A,2023PhRvX..13a1048A,2025arXiv250818083T}, subsequent analyses have shown that its robustness is model dependent~\cite{2024ApJ...962...69F,2025PhRvD.111f1305H,2025arXiv251122093S,2025ApJ...994L..52T} and that it could arise as a statistical fluctuation when compared against mock catalogs~\cite{2023ApJ...955..107F}. Our bootstrap analysis reinforces this more cautious interpretation.  
Likewise, tentative features currently identified in the spin-orientation distribution (e.g.~\cite{2025arXiv251215873S}) become considerably more uncertain once catalog variance is included, casting doubt on their robustness~\cite{2025PhRvD.112h3015V}.

We argue that incorporating the catalog variance is a necessary step in GW population studies. Accounting for the finite-size of the catalog with methods such as the one presented here is crucial to prevent overinterpreting features and correlations in the reconstructed GW population that may not persist as the catalog grows. %

\vspace{\baselineskip}
\section{Acknowledgements}
We thank Tristan Bruel, Stephen Green, Costantino Pacilio, Rodrigo Tenorio, and Alexandre Toubiana for discussions.
A.C. and D.G. are supported by 
ERC Starting Grant No.~945155--GWmining, 
Cariplo Foundation Grant No.~2021-0555, 
MUR PRIN Grant No.~2022-Z9X4XS, 
Italian-French University (UIF/UFI) Grant No.~2025-C3-386,
MUR Grant ``Progetto Dipartimenti di Eccellenza 2023-2027'' (BiCoQ),
the ICSC National Research Centre funded by NextGenerationEU,
and the INFN TEONGRAV initiative.
D.G. is supported by MSCA Fellowship No.~101149270--ProtoBH and MUR Young Researchers Grant No. SOE2024-0000125.
M.M. is supported by a Research Fellowship from the Royal Commission for the Exhibition of 1851.
Computational work was performed at CINECA with allocations through INFN and the University of Milano-Bicocca.
This material is based upon work supported by NSF's LIGO Laboratory which is a major facility fully funded by the National Science Foundation.

\bibliography{popvariance}

\onecolumngrid
\vspace{0.5cm}
\begin{center}
{\bf\large End Matter}
\end{center}
\vspace{0.2cm}
\twocolumngrid

\begin{figure*}
    \centering
    \includegraphics[width=0.97\textwidth]{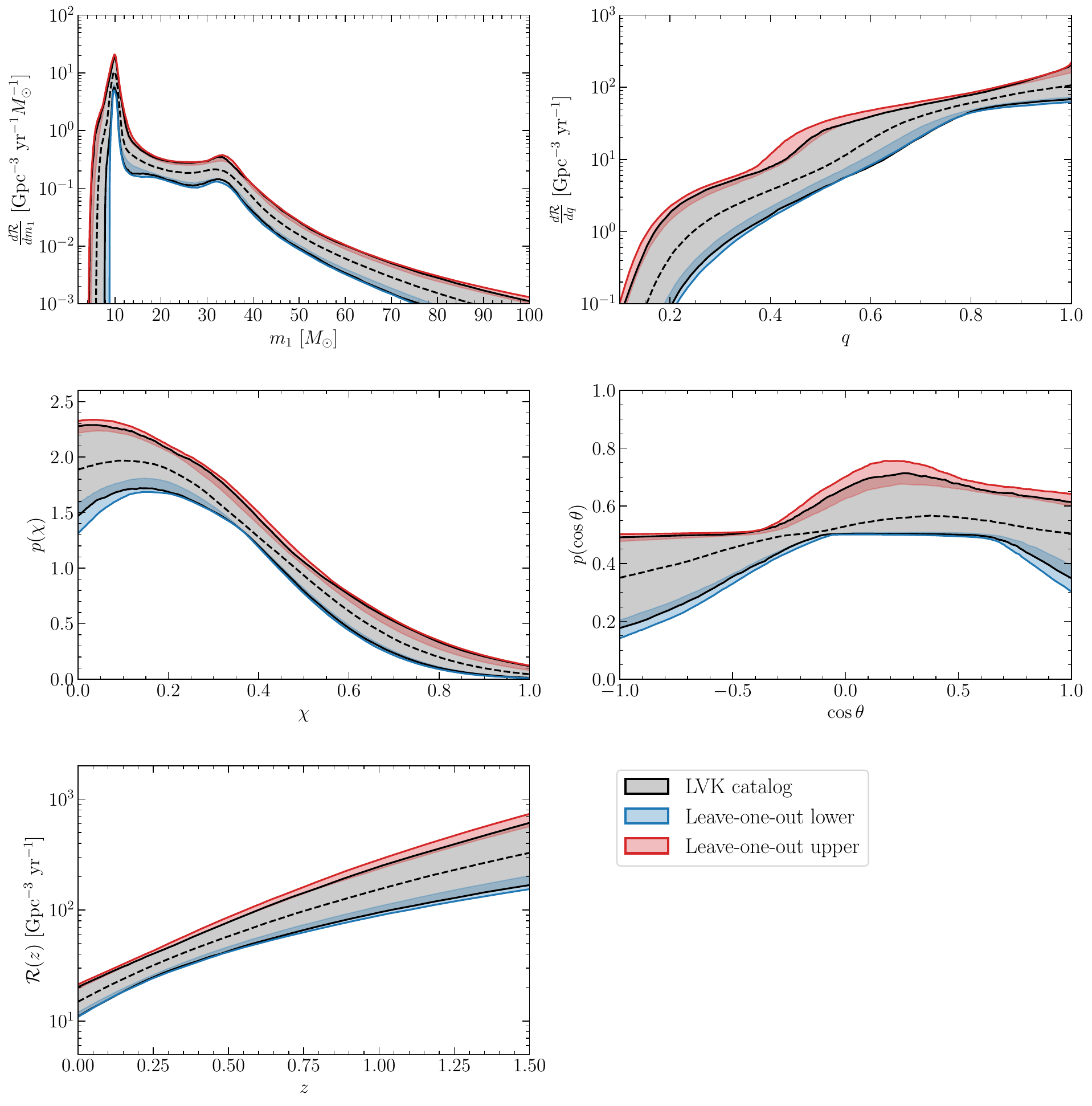}
    \caption{As in Fig.~\ref{fig:catalog_variance}, but using leave-one-out resampling instead of bootstraps.
    }
    \label{fig:leave_one_out_plots}
\end{figure*}

\begin{figure*}
    \centering
    \includegraphics[width=0.97\textwidth]{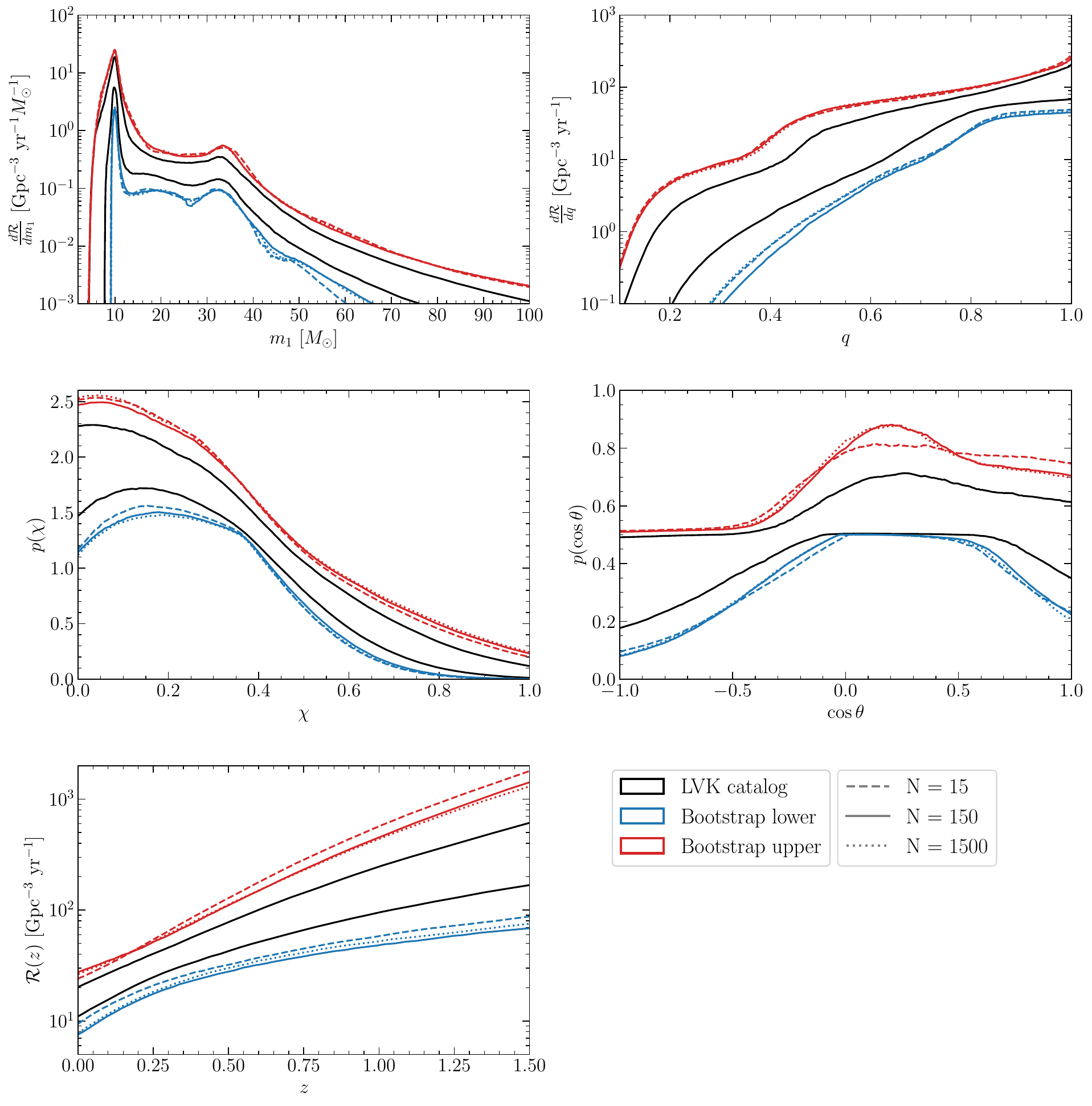}
    \caption{As in Fig.~\ref{fig:catalog_variance}, but showing the effect of varying the number of bootstrap segments. We consider $N=15$ (dashed), $N=150$ (solid; the default value used in the main body of the paper), and $N=1500$ (dotted).
    }
    \label{fig:segments_variation}
\end{figure*}

\section{Leave-one-out analysis}  
Leave-one-out resampling (or jackknifing) is a statistical approach alternative to bootstrapping for computing the variance of an estimator. Given $N$ datapoints, one recomputes the estimator $N$ times, each time excluding one datapoint. The spread of the resulting distribution is then used to estimate the variance of the estimator computed with all $N$ datapoints. 
We apply this technique to the problem of GW populations, using the same $N=150$ time segments described in the main body of the paper. The analysis proceeds as outlined above, with the only difference being that the total observing time entering the rate calculation must be corrected by a factor $(N-1)/N$.

Results are presented in Fig.~\ref{fig:leave_one_out_plots}, which shows that a catalog variance lower than that obtained via bootstraps in Fig.~\ref{fig:catalog_variance}. This is not surprising, as leave-one-out methods are known to generically underestimate an estimator's variance~\cite{tibshirani1993introduction,1982jbor.book.....E,shao2012jackknife}. More formally, one can show that leave-one-out approximations reproduce bootstrap results only for sufficiently simple estimators, which is very likely not the case here, as each of the curves in Fig.~\ref{fig:leave_one_out_plots} is the result of a multi-step inference procedure. Furthermore, leave-one-out approaches have known pitfalls when applied to order statistics~\cite{2020sdmm.book.....I}, such as the quantiles shown here. For these reasons, we believe the bootstrap results presented in the main body of the paper provide a more faithful representation of the GWTC-4 catalog variance.

Tangentially, we note that a leave-one-out analysis cannot be performed solely on the detected events. That is, one cannot exclude an event from the GW catalog and simply rerun the population pipeline, as this would effectively increase the time during which the detector has not observed GWs, thereby impacting the estimation of selection effects~\cite{2022ApJ...926...34E} (in practice, this is equivalent to performing a population analysis for a different detector). Unfortunately, such practice is common in GW astronomy.

\section{Impact of the number of time segments} \label{Nchunks}
In Fig.~\ref{fig:segments_variation}, we reproduce the main results of this study for different choices of the number of bootstrap segments, $N=15$, 150 (default), and 1500. In all cases, we perform 700 bootstrap realizations. The results are largely unchanged across these choices. %

\end{document}